\documentclass[aps,prb,twocolumn,showpacs,floats]{revtex4}
\usepackage{graphics}
\usepackage{amssymb}
\begin{document}

\title{Low temperature quasiparticle transport in a $d$-wave
superconductor with coexisting charge order}
\author{Adam C. Durst$^1$ and Subir Sachdev$^2$}
\affiliation{$^1$Department of Physics and Astronomy, Stony Brook
University, Stony Brook, NY 11794-3800 \\ $^2$Department of
Physics, Harvard University, Cambridge, MA 02138}
\date{October 21, 2008}

\begin{abstract}
In light of the evidence that charge order coexists with $d$-wave
superconductivity in the underdoped cuprate superconductors, we
investigate the manner in which such charge order will influence
the quasiparticle excitations of the system and, in particular,
the low-temperature transport of heat by those quasiparticles.  We
consider a d-wave superconductor in which the superconductivity
coexists with charge density wave order of wave vector
$(\pi/a,0)$. While the nodes of the quasiparticle energy spectrum
survive the onset of charge order, there exists a critical value
of the charge density wave order parameter beyond which the
quasiparticle spectrum becomes fully gapped.  We perform a linear
response Kubo formula calculation of thermal conductivity in the
low temperature (universal) limit. Results reveal the dependence
of thermal transport on increasing charge order up to the critical
value at which the quasiparticle spectrum becomes fully gapped and
thermal conductivity vanishes. In addition to numerical results,
closed-form expressions are obtained in the clean limit for the
special case of isotropic Dirac nodes. Signatures of the influence
of charge order on low-temperature thermal transport are
identified.
\end{abstract}


\maketitle

\section{Introduction}
\label{sec:intro} The low energy quasiparticle excitations of the
$d$-wave superconducting phase of the high-$T_c$ cuprate
superconductors are massless anisotropic Dirac fermions.
\cite{lee97}  These Dirac quasiparticles are easily excited in the
vicinity of the four nodes, the four points on the two-dimensional
Fermi surface where the superconducting order parameter vanishes.
The dominant carriers of heat at low temperature, quasiparticles
are efficiently probed via low temperature thermal conductivity
measurements, which have been performed extensively over the past
decade.  Theory \cite{lee93,hir93,hir94,hir96,gra96,sen98,dur00}
has shown that the massless Dirac energy spectrum yields a low
temperature limit where thermal conductivity is remarkably
independent of disorder for small impurity density. In this limit,
known as the universal limit, thermal conductivity per CuO$_2$
plane depends only on fundamental constants and the ratio of the
Fermi velocity, $v_F$, to the $k$-space slope of the
superconducting order parameter at the nodal points, $v_\Delta$.
Experiments
\cite{tai97,chi99,chi00,pro02,sut03,hil04,sun04,sut05,haw07,sun08}
have demonstrated this disorder-independence and used this result
to extract the anisotropy ratio, $\alpha \equiv v_F / v_\Delta$,
from low-temperature thermal transport data.

Over the past few years, there has been a significant effort to
grow and measure high quality cuprate samples in the underdoped
regime of the superconducting phase, as well as the pseudogap
phase that results from underdoping even further. Several
experimental groups have used high-resolution scanning tunneling
microscopy
\cite{hof02a,hof02b,ver04,han04,mis04,mce05,koh07,boy07,han07,pas08,wis08,koh08}
to examine the electronic states of the underdoped cuprates at the
atomic scale. These experiments, amongst others \cite{kiv03}, have
provided evidence that charge order coexists with $d$-wave
superconductivity (dSC) in these materials. Furthermore, it has
been shown theoretically \cite{par01,gra01,voj00} that coexisting
charge order can significantly affect the quasiparticle spectrum
of the superconductor, leading the system to become fully gapped
for charge order of sufficient magnitude. If the quasiparticles
are fully gapped (no nodes), the dominant carriers of heat at low
temperature are frozen out, which should have a dramatic effect on
the universal-limit thermal conductivity.  The purpose of our
current analysis is to study the nature of this effect.

We consider a particularly simple form of charge order, a
conventional $s$-wave charge density wave (CDW) with a
$k$-independent order parameter and a wave vector ${\bf
Q}=(\pi/a,0)$ that doubles the unit cell.  While the charge order
in the underdoped cuprates may be of a more complex type, this
simple model provides a place for us to start studying,
phenomenologically, the effect of charge order on thermal
transport in a $d$-wave superconductor.  Furthermore, since the
experimentally observed \cite{han04} CDW has a wave vector close
to $(\pi/2a,0)$, it will generically have a second harmonic near
$(\pi/a,0)$.  This harmonic can couple efficiently to the nodal
quasiparticles because its wave vector nearly spans the separation
between the nodes. \cite{ber08}  The calculations presented in
this paper can then be viewed as applying to this second harmonic.

While the charge order in the cuprates may turn on with
underdoping, we simply add a CDW term to the dSC Hamiltonian and
turn on the charge order by hand, by increasing the magnitude of
the CDW order parameter.  We then calculate the universal limit
thermal conductivity of the combined system, evaluating the effect
of coexisting charge order on thermal transport.  Our goal is to
identify signatures of the onset of charge order which may be
observed with underdoping in low-temperature thermal conductivity
measurements of the underdoped cuprates.  Evidence of the
breakdown of universal thermal conductivity at low doping,
possibly due to the onset of charge order, has already been seen
experimentally \cite{and04,sun05,sun06,haw03}, and our results
might therefore shed light on these studies.

We begin in Sec.~\ref{sec:setup} by writing down the combined
Hamiltonian, calculating the resulting energy spectrum, and
discussing the charge-order-induced transition whereby the
spectrum can become fully gapped. In Sec.~\ref{sec:transportcalc},
we calculate the Green's function and thermal current operator,
define our model of disorder, and use a diagrammatic Kubo formula
approach to obtain an integral form for the thermal conductivity
tensor.  For the special case of a clean system (no disorder) with
isotropic nodes ($v_F = v_\Delta$) the remaining $k$-space
integration can be performed analytically. This case is considered
in Sec.~\ref{sec:analytical} where a closed-form solution is
obtained for the thermal conductivity tensor as a function of the
magnitude of the charge order.  The more general case of nonzero
disorder and anisotropic nodes is considered in
Sec.~\ref{sec:numerical} via a numerical computation, and the
effect of disorder and nodal anisotropy is discussed. Conclusions
are presented in Sec.~\ref{sec:conclusions}.

\section{Coexisting $d$SC and CDW Order}
\label{sec:setup}

\subsection{Hamiltonian}
\label{ssec:Hamiltonian}
Following Ref.~\onlinecite{par01}, we
consider a model Hamiltonian for a $d$-wave superconductor with
coexisting charge order:
\begin{equation}
H = H_0 + H_{\rm dSC} + H_{\rm CDW}
\end{equation}
\begin{equation}
H_0 = \sum_{k\sigma} \epsilon_{k} c_{k\sigma}^{\dagger}
c_{k\sigma}
\end{equation}
\begin{equation}
H_{\rm dSC} = \sum_{k} \Delta_{k} \left( c_{k\uparrow}^{\dagger}
c_{-k\downarrow}^{\dagger} + c_{-k\downarrow} c_{k\uparrow}
\right)
\end{equation}
\begin{equation}
H_{\rm CDW} = \sum_{k\sigma} \psi\ c_{k\sigma}^{\dagger}
c_{k+Q\sigma}
\end{equation}
Momenta are summed over the Brillouin zone of a two-dimensional
square lattice of lattice constant $a$.  $H_0 + H_{\rm dSC}$ is
the mean-field BCS Hamiltonian for electron dispersion
$\epsilon_k$ and superconducting order parameter $\Delta_k$, which
is taken to have $d$-wave symmetry (for example,
$\Delta_{k}=\Delta_{0}(\cos k_x a - \cos k_y a)/2$). $H_{\rm CDW}$
denotes a charge density wave of wave vector ${\bf Q}$ with CDW
order parameter $\psi$.  While it is possible to consider
density-wave states of nonzero angular momentum \cite{nay00} by
taking $\psi$ to be complex and $k$-dependent, we shall focus here
on the effect of a conventional $s$-wave CDW corresponding to a
site-centered charge modulation in the $x$-direction of wavelength
twice the lattice constant.  That is, we take $\psi$ to be a real,
$k$-independent parameter and set ${\bf Q}=(\pi/a, 0)$.

\begin{figure}
\centerline{\resizebox{3.25in}{!}{\includegraphics{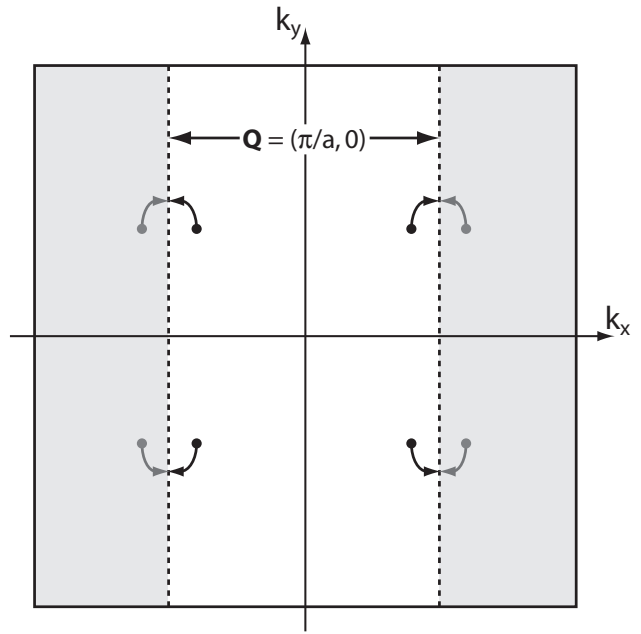}}}
\caption{Charge order of wave vector ${\bf Q} = (\pi/a,0)$ doubles
the unit cell and thereby halves the Brillouin zone. With
increasing charge density wave order parameter, $\psi$, the nodes
of the energy spectrum, and their images in the second reduced
Brillouin zone (shaded), approach the reduced Brillouin zone edges
(dotted), colliding for $\psi = \psi_c$, beyond which the spectrum
is fully gapped.}
\label{fig:BZ}
\end{figure}

The charge density wave has the effect of doubling the unit cell
and therefore halving the effective Brillouin zone, as shown in
Fig.~\ref{fig:BZ}.  By defining a four-component extended-Nambu
vector
\begin{equation}
\Psi_{k}^{\dagger} = \left[
\begin{array}{cccc} c^{\dagger}_{k\uparrow}, & c_{-k\downarrow}, &
c^{\dagger}_{k+Q\uparrow}, & c_{-k-Q\downarrow} \end{array}
\right]
\label{eq:basis}
\end{equation}
consisting of particle and hole operators at ${\bf k}$ and ${\bf
k}+{\bf Q}$, we can express the Hamiltonian in a compact $4 \times
4$ matrix notation:
\begin{equation}
H = \sum_{k}^{\prime} \Psi_{k}^{\dagger} H_{k} \Psi_{k}
\end{equation}
where
\begin{equation}
H_k = \left[ \begin{array}{cccc}
\epsilon_1 & \Delta_1 & \psi & 0 \\
\Delta_1 & -\epsilon_1 & 0 & -\psi \\
\psi & 0 & \epsilon_2 & \Delta_2 \\
0 & -\psi & \Delta_2 & -\epsilon_2 \end{array} \right]
\label{eq:Hk}
\end{equation}
and subscript $1$ denotes ${\bf k}$ and subscript $2$ denotes
${\bf k}+{\bf Q}$.  The prime indicates that the momentum sum is
restricted to the reduced Brillouin zone.  Note that the
upper-left and lower-right $2 \times 2$ blocks of $H_{k}$ are
simply the Nambu space Hamiltonian at ${\bf k}$ and ${\bf k}+{\bf
Q}$ respectively. The CDW order parameter couples these two
sectors.

\subsection{Energy Spectrum and Nodal Collision}
\label{ssec:spectrum}
The energy spectrum of the fermionic
excitations of this system of coexisting dSC and CDW order is
obtained by solving for the (positive) eigenvalues of $H_k$. Doing
so, we find that
\begin{eqnarray}
\lefteqn{E_k = \frac{1}{2} \Big\{ \left( \epsilon_1^2 + \Delta_1^2
+ \epsilon_2^2 + \Delta_2^2 + 2\psi^2 \right)} \\ \nonumber && \pm
\Big[ \left( \epsilon_1^2 + \Delta_1^2 - \epsilon_2^2 - \Delta_2^2
\right)^2 \\ \nonumber && + 4\psi^2 \left( (\epsilon_1 +
\epsilon_2)^2 + (\Delta_1 - \Delta_2)^2 \right) \Big]^{1/2}
\Big\}^{1/2}
\end{eqnarray}
For $\psi=0$, these solutions reduce to the energy spectra of the
quasiparticle excitations of the $d$-wave superconductor, $E_k^0$
and $E_{k+Q}^0$, where $E_k^0=\sqrt{\epsilon_k^2+\Delta_k^2}$. By
construction, the quasiparticle energies drop to zero at four
points in the Brillouin zone, the intersection of the Fermi
surface with the lines $k_x=\pm k_y$.  These are the nodal points,
or nodes, of the $d$-wave superconductor.  To model the situation
in the cuprates, the nodes are taken to be a distance $k_F$ from
the origin, inside of the $(\pm\pi/2a,\pm\pi/2a)$ points by a
small distance $k_0$ where $k_0 \equiv \pi/\sqrt{2}a - k_F \ll
k_F$.

As the charge density wave is turned on, the nodal structure of
the excitation spectrum initially survives, since the CDW
wavevector, ${\bf Q}$, is not commensurate with the internodal
distance. \cite{par01,ber08}  With increasing $\psi$, the nodes
move toward the reduced Brillouin zone edge along the trajectory
sketched in Fig.~\ref{fig:BZ}.  Also plotted in this figure is the
trajectory of the image of each node translated by ${\bf Q}$ into
the second reduced Brillouin zone.  At a critical value of the CDW
order parameter, $\psi=\psi_c$, the nodes collide at the reduced
Brillouin zone edge and the energy spectrum becomes fully gapped.
For $\psi > \psi_c$, the minimum values of the excitation spectra
are nonzero.  Hence, the nodes have vanished.

\begin{figure}
\centerline{\resizebox{3.25in}{!}{\includegraphics{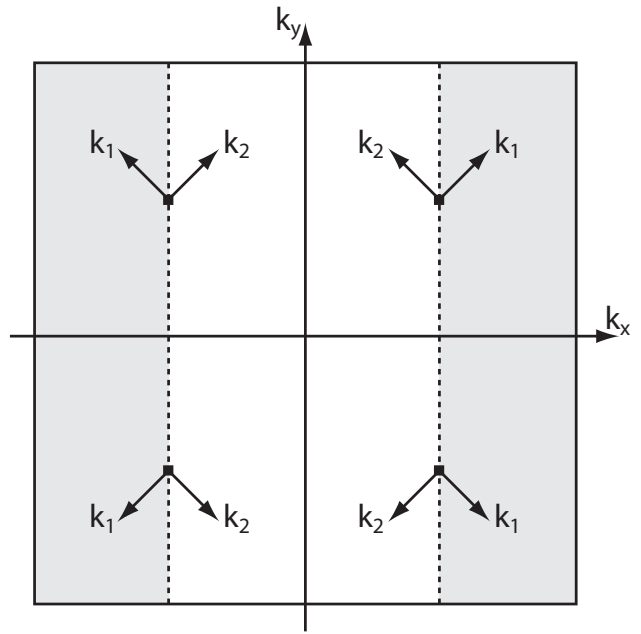}}}
\caption{Local coordinates, $k_1$ and $k_2$, defined about each of
the four nodal collision points, ${\bf k}_c = (\pm \pi/2a, \pm
\pi/2a)$.  The $k_1$-axes, perpendicular to the Fermi surface,
define the direction of increasing electron dispersion,
$\epsilon_k$.  The $k_2$-axes, parallel to the Fermi surface,
define the direction of increasing superconducting order
parameter, $\Delta_k$.}
\label{fig:BZcoordinates}
\end{figure}

To determine the points in $k$-space at which this nodal collision
occurs, we need only solve for the points at which a zero of $E_k$
coincides with a reduced Brillouin zone edge.  We define the
$\psi$ for which this occurs to be $\psi_c$.  For example, node
\#1 (located in the upper-right quadrant) will collide somewhere
along the reduced zone boundary at $k_x=\pi/2a$.  Setting
$E(k_x=\pi/2a,k_y)=0$ and noting that both $\epsilon_k$ and
$\Delta_k$ are even functions of $k_x$, we find that the collision
point must satisfy $\epsilon_1=\epsilon_2=\psi_c$ and
$\Delta_1=\Delta_2=0$.  Near node \#1, the latter condition yields
$k_x=k_y$, so the collision point is ${\bf k}_c=(\pi/2a,\pi/2a)$.
Equivalent arguments for each of the four quadrants reveal that
the four collision points are located at $(\pm \pi/2a,\pm
\pi/2a)$.  Defining local coordinates $k_1$ and $k_2$ about each
of the collision points, as shown in Fig.~\ref{fig:BZcoordinates},
we can write
\begin{eqnarray}
\epsilon_1 = v_F(k_0 + k_1) && \Delta_1 = v_\Delta k_2 \nonumber \\
\epsilon_2 = v_F(k_0 + k_2) && \Delta_2 = v_\Delta k_1
\end{eqnarray}
where $v_F$ is the Fermi velocity and $v_\Delta$ is the slope of
the gap at the node.  Note that in writing these linear relations,
we have assumed that $k_0$ is small enough that the spectrum of
the $d$-wave superconductor is still linear in the vicinity of the
collision points.  At the collision points ($k_1=k_2=0$),
$\epsilon_1 = \epsilon_2 = v_F k_0$, which requires that $\psi_c =
v_F k_0$.  Switching to scaled coordinates, $p_1 \equiv \sqrt{v_F
v_\Delta} k_1$ and $p_2 \equiv \sqrt{v_F v_\Delta} k_2$, yields
\begin{eqnarray}
\epsilon_1 = \psi_c + \sqrt{\alpha} p_1 && \Delta_1 = p_2 / \sqrt{\alpha} \nonumber \\
\epsilon_2 = \psi_c + \sqrt{\alpha} p_2 && \Delta_2 = p_1 /
\sqrt{\alpha}
\label{eq:epsdelta}
\end{eqnarray}
where $\alpha \equiv v_F / v_\Delta$.  This notation provides a
convenient framework with which to proceed with the thermal
transport calculation.

\section{Transport Calculation}
\label{sec:transportcalc} Given the Hamiltonian defined by
Eqs.~(\ref{eq:Hk}) and (\ref{eq:epsdelta}), we can calculate the
thermal conductivity, and its dependence on the charge density
wave order parameter, via Kubo formula.

\subsection{Green's Function}
\label{ssec:Greenfunc}
We begin by computing the Matsubara Green's
function. In the extended-Nambu basis of Eq.~(\ref{eq:basis}), the
bare Green's function is a $4 \times 4$ matrix obtained through
inversion of the Hamiltonian
\begin{equation}
G^0(k,i\omega) = \left[ i\omega - H_k \right]^{-1}
\label{eq:G0def}
\end{equation}
It takes the form
\begin{equation}
G^0(k,\omega) = \frac{1}{G_{\rm den}} \left[
\begin{array}{cc}
G_a & G_b \\
G_c & G_d
\end{array}
\right]
\label{eq:G0}
\end{equation}
\begin{eqnarray}
G_a &=& ((i\omega)^2 - \epsilon_2^2 - \Delta_2^2) [i\omega +
\epsilon_1 \tau_3 + \Delta_1 \tau_1] \nonumber \\
&& - \psi^2 [i\omega - \epsilon_2 \tau_3 + \Delta_2 \tau_1]
\label{eq:Ga}
\end{eqnarray}
\begin{eqnarray}
G_b &=& \psi \big[ i\omega (\epsilon_1+\epsilon_2) + ((i\omega)^2
+ \epsilon_1\epsilon_2 - \Delta_1\Delta_2 - \psi^2)\tau_3 \nonumber \\
&& + (\epsilon_1\Delta_2 + \epsilon_2\Delta_1)\tau_1 -
i\omega(\Delta_1-\Delta_2)(i\tau_2) \big]
\label{eq:Gb}
\end{eqnarray}
\begin{eqnarray}
G_c &=& \psi \big[ i\omega (\epsilon_1+\epsilon_2) + ((i\omega)^2
+ \epsilon_1\epsilon_2 - \Delta_1\Delta_2 - \psi^2)\tau_3 \nonumber \\
&& + (\epsilon_1\Delta_2 + \epsilon_2\Delta_1)\tau_1 +
i\omega(\Delta_1-\Delta_2)(i\tau_2) \big]
\label{eq:Gc}
\end{eqnarray}
\begin{eqnarray}
G_d &=& ((i\omega)^2 - \epsilon_1^2 - \Delta_1^2) [i\omega +
\epsilon_2 \tau_3 + \Delta_2 \tau_1] \nonumber \\
&& - \psi^2 [i\omega - \epsilon_1 \tau_3 + \Delta_1 \tau_1]
\label{eq:Gd}
\end{eqnarray}
\begin{eqnarray}
G_{\rm den} &=& (\epsilon_1^2+\Delta_1^2+\psi^2-(i\omega)^2)
(\epsilon_2^2+\Delta_2^2+\psi^2-(i\omega)^2) \nonumber \\
&& -\psi^2((\epsilon_1+\epsilon_2)^2+(\Delta_1-\Delta_2)^2)
\label{eq:Gden}
\end{eqnarray}
where $G_{\rm den}$ is a scalar and $G_a$, $G_b$, $G_c$, and $G_d$
are $2 \times 2$ matrices expressed in terms of
particle-hole-space Pauli matrices, $\tau_i$.

In the presence of disorder, we must include the impurity
contribution to the self-energy via Dyson's equation
\begin{equation}
G^{-1} = \left. G^0 \right.^{-1} - \Sigma
\label{eq:Dyson}
\end{equation}
The self-energy, $\Sigma$, is a $4 \times 4$ matrix in the
extended-Nambu basis, but for simplicity, we consider here only
the scalar term
\begin{equation}
\Sigma = \Sigma(i\omega) \openone
\label{eq:scalarSigma}
\end{equation}
and postpone discussion of the effects of off-diagonal self-energy
terms to a separate publication \cite{sch08}.  Then the dressed
Matsubara Green's function is simply
\begin{equation}
G(k,i\omega) = \left[ (i\omega-\Sigma(i\omega))\openone - H_k
\right]^{-1} = G^0(k,i\omega-\Sigma(i\omega))
\label{eq:Gdresseddef}
\end{equation}
For our calculation of the zero-temperature thermal conductivity,
we will require only the imaginary part of the zero-frequency
retarded Green's function, $\mbox{Im}\,G^R(k,\omega \rightarrow
0)$. Continuing $i\omega \rightarrow \omega + i\delta$ and taking
the $\omega \rightarrow 0$ limit, the zero-frequency retarded
self-energy is just a negative imaginary constant, $-i\Gamma_0$,
and we find that
\begin{equation}
\mbox{Im}\,G^R(k,\omega \rightarrow 0) = \frac{1}{G_{\rm den}}
\left[
\begin{array}{cc}
G^{''}_a & G^{''}_b \\
G^{''}_c & G^{''}_d
\end{array}
\right]
\label{eq:ImG}
\end{equation}
\begin{equation}
G^{''}_a = -\Gamma_0 (\Gamma_0^2 + \psi^2 + \epsilon_2^2 +
\Delta_2^2)
\label{eq:ImGa}
\end{equation}
\begin{equation}
G^{''}_b = \psi \Gamma_0 \left[ (\epsilon_1+\epsilon_2) -
(\Delta_1 - \Delta_2)(i\tau_2) \right]
\label{eq:ImGb}
\end{equation}
\begin{equation}
G^{''}_c = \psi \Gamma_0 \left[ (\epsilon_1+\epsilon_2) +
(\Delta_1 - \Delta_2)(i\tau_2) \right]
\label{eq:ImGc}
\end{equation}
\begin{equation}
G^{''}_d = -\Gamma_0 (\Gamma_0^2 + \psi^2 + \epsilon_1^2 +
\Delta_1^2)
\label{eq:ImGd}
\end{equation}
\begin{eqnarray}
G_{\rm den} &=& (\Gamma_0^2+\psi^2+\epsilon_1^2+\Delta_1^2)
(\Gamma_0^2+\psi^2+\epsilon_2^2+\Delta_2^2) \nonumber \\
&& -\psi^2((\epsilon_1+\epsilon_2)^2+(\Delta_1-\Delta_2)^2)
\label{eq:ImGden}
\end{eqnarray}
where $\Gamma_0$ is the zero-frequency impurity scattering rate
(the impurity-induced broadening of the spectral function).

\subsection{Current Operator}
\label{ssec:currentoperator}
Next we must calculate the
quasiparticle current operator for this system of coexisting
$d$-wave superconductor and charge order. We note that
quasiparticles carry a well-defined heat and spin. Thus, where a
quasiparticle goes, so goes its heat and spin. Though the quantity
we require is the thermal current, we will proceed by calculating
the spin current operator (which is technically simpler) obtaining
the thermal current operator by correspondence.

The spin current operator, ${\bf j}^s$, is obtained via continuity
with the spin density operator, $\rho^s$.
\begin{equation}
-{\bf \nabla} \cdot {\bf j}^s = \dot{\rho^s} = \frac{1}{i} \left[
\rho^s, H \right]
\label{eq:continuity}
\end{equation}
Fourier transforming and taking the zero-wavevector limit yields a
recipe for calculating ${\bf j}^s_{q=0}$, which is the operator we
will need for the transport calculation.
\begin{equation}
{\bf q} \cdot {\bf j}^s_0 = \lim_{q \rightarrow 0} \left[
\rho^s_q, H \right]
\label{eq:continuityq}
\end{equation}
Defining and re-expressing the spin density operator in various
forms, we note that
\begin{eqnarray}
\rho^s_q &\equiv& \sum_{k^{'}\sigma} S_\sigma
c^{\dagger}_{k^{'}\sigma} c_{k^{'}+q \sigma} \nonumber \\
&=& s \sum_{k^{'}}^{\prime} \Psi^\dagger_{k^{'}} \Psi_{k^{'}+q}
\nonumber \\
&=& s \sum_{k^{'}}^{\prime} \Big( c^\dagger_{k^{'}\uparrow}
c_{k^{'}+q\uparrow} + c_{-k^{'}\downarrow}
c^\dagger_{-k^{'}-q\downarrow} \nonumber \\
&& + d^\dagger_{k^{'}\uparrow} d_{k^{'}+q\uparrow} +
d_{-k^{'}\downarrow} d^\dagger_{-k^{'}-q\downarrow} \Big)
\label{eq:spindensity}
\end{eqnarray}
where $S_\sigma = \pm s$, $s=1/2$, $d_{k\sigma} \equiv
c_{k+Q\sigma}$, $\Psi_k$ is the four-component extended-Nambu
vector defined in Eq.~(\ref{eq:basis}), and the prime restricts
the wave vector sum to the reduced Brillouin zone. In the same
notation, the Hamiltonian takes the form
\begin{eqnarray}
&H& = \sum_k^\prime \Psi^\dagger_k H_k \Psi_k \nonumber \\
&=& \sum_k^\prime \Big[ \epsilon_k (c^\dagger_{k\uparrow}
c_{k\uparrow} - c_{-k\downarrow} c^\dagger_{-k\downarrow}) +
\Delta_k (c^\dagger_{k\uparrow} c^\dagger_{-k\downarrow} +
c_{-k\downarrow} c_{k\uparrow}) \nonumber \\
&+& \epsilon_{k+Q} (d^\dagger_{k\uparrow} d_{k\uparrow} -
d_{-k\downarrow} d^\dagger_{-k\downarrow}) + \Delta_{k+Q}
(d^\dagger_{k\uparrow} d^\dagger_{-k\downarrow} +
d_{-k\downarrow} d_{k\uparrow}) \nonumber \\
&+& \psi (c^\dagger_{k\uparrow} d_{k\uparrow} - c_{-k\downarrow}
d^\dagger_{-k\downarrow} + d^\dagger_{k\uparrow} c_{k\uparrow} -
d_{-k\downarrow} c^\dagger_{-k\downarrow}) \Big]
\label{eq:HamWrittenOut}
\end{eqnarray}
Using fermion anticommutation relations to evaluate the commutator
in Eq.~(\ref{eq:continuityq}), we find that
\begin{equation}
{\bf j}^s_0 = s \sum_k^\prime \Psi^\dagger_k \left[
\begin{array}{cc}
{\bf v}_{Fk} \tau_3 + {\bf v}_{\Delta k} \tau_1 & {\bf v}_{\psi k}
\tau_3 \\
{\bf v}_{\psi k} \tau_3 & {\bf v}_{F\,k+Q} \tau_3 + {\bf
v}_{\Delta\, k+Q} \tau_1
\end{array}
\right] \Psi_{k+Q}
\label{eq:js0array}
\end{equation}
where ${\bf v}_{F k} \equiv \partial \epsilon_k / \partial {\bf
k}$, ${\bf v}_{\Delta k} \equiv \partial \Delta_k /
\partial {\bf k}$, and ${\bf v}_{\psi k} \equiv
\partial \psi / \partial {\bf k}$.  For the case we consider,
$\psi$ is $k$-independent, so ${\bf v}_{\psi k}$ is precisely zero
and the spin current operator is block diagonal in the
extended-Nambu basis.

In the vicinity of each of the four collision points (the regions
we will always be considering), ${\bf v}_{Fk}$ points along the
locally-defined $k_1$-direction and ${\bf v}_{\Delta k}$ points
along the locally-defined $k_2$-direction, as shown in
Fig.~\ref{fig:BZcoordinates}.  Therefore, shifting by wave vector
${\bf Q} = (\pi/a,0)$ from ${\bf k}$ to ${\bf k}+{\bf Q}$ flips
the sign of the $x$-component of each velocity while preserving
the $y$-component.  That is, the components satisfy $v_{F\,k+Q}^i
= \eta_i v_{Fk}^i$ and $v_{\Delta\,k+Q}^i = \eta_i v_{\Delta k}^i$
for
\begin{equation}
\eta_i \equiv \left\{
\begin{array}{c}
-1 \;\;\mbox{for}\;\; i=x \\
+1 \;\;\mbox{for}\;\; i=y
\end{array}
\right\}
\label{eq:eta}
\end{equation}
So we can write
\begin{equation}
{\bf j}^s_0 = s \sum_k^\prime \Psi^\dagger_k \left[ {\bf v}_{MF} +
{\bf v}_{M\Delta} \right] \Psi_{k+Q}
\label{eq:js0vv}
\end{equation}
where
\begin{equation}
{\bf v}_{MF} \equiv v_{Fk}^x M_3^x \hat{\bf x} + v_{Fk}^y M_3^y
\hat{\bf y}
\label{eq:vMF}
\end{equation}
\begin{equation}
{\bf v}_{M\Delta} \equiv v_{\Delta k}^x M_1^x \hat{\bf x} +
v_{\Delta k}^y M_1^y \hat{\bf y}
\label{eq:vMD}
\end{equation}
\begin{equation}
M_3^i \equiv \left(
\begin{array}{cc}
\tau_3 & 0 \\
0 & \eta_i \tau_3
\end{array}
\right) \;\;\;\;\;\;
M_1^i \equiv \left(
\begin{array}{cc}
\tau_1 & 0 \\
0 & \eta_i \tau_1
\end{array}
\right)
\label{eq:M3M1}
\end{equation}

Finally, we note that since the same quasiparticles that carry the
spin also carry the heat, the thermal current operator, ${\bf
j}^{\kappa}$, will have the same structure as the spin current
operator.  In the zero-wavevector, zero-frequency limit that we
will require,
\begin{equation}
{\bf j}^\kappa_0 = \lim_{q,\Omega \rightarrow 0}
\sum_{k\omega}^\prime (\omega + \frac{\Omega}{2})
\Psi^\dagger_{k,\omega} \left[ {\bf v}_{MF} + {\bf v}_{M\Delta}
\right] \Psi_{k+Q,\omega+\Omega}
\label{eq:heatcurrent}
\end{equation}

\subsection{Thermal Conductivity}
\label{ssec:thermalconductivity} Given the Green's function,
thermal current operator, and coordinate system defined in the
previous sections, we can calculate the thermal conductivity via
the Kubo formula \cite{mah90}
\begin{equation}
\frac{\tensor{\kappa}(T)}{T} = -\lim_{\Omega \rightarrow 0}
\frac{\mbox{Im}\,\tensor{\Pi}_\kappa^R(\Omega)}{T^2 \Omega}
\label{eq:kubo}
\end{equation}
where the retarded current-current correlation function is
obtained from the Matsubara function via analytic continuation.
\begin{equation}
\tensor{\Pi}_\kappa^R(\Omega) = \tensor{\Pi}_\kappa(i\Omega
\rightarrow \Omega + i\delta)
\label{eq:Picontinue}
\end{equation}
In what follows, we neglect vertex corrections, calculating the
bare bubble current-current correlation function using the
Matsubara formalism \cite{mah90}.  It has been shown previously
\cite{dur00} that vertex corrections are negligible for the
$d$-wave superconductor case (without charge order) and the
contribution of vertex corrections to the present case will be
considered in the a separate paper \cite{sch08}.

\begin{figure}
\centerline{\resizebox{2in}{!}{\includegraphics{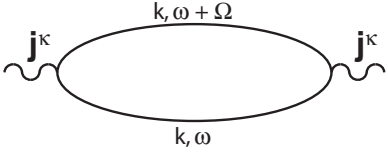}}}
\caption{Feynman diagram depicting the bare bubble thermal
current-current correlation function,
$\tensor{\Pi}_\kappa(i\Omega)$.  The thermal current operator sits
on each vertex and each propagator denotes a Green's function
dressed with disorder self-energy.}
\label{fig:bubble}
\end{figure}

Evaluating the bare bubble Feynman diagram shown in
Fig.~\ref{fig:bubble} yields
\begin{eqnarray}
&& \tensor{\Pi}_\kappa (i\Omega) = \frac{1}{\beta} \sum_{i\omega}
\sum_k^\prime (i\omega + \frac{i\Omega}{2})^2 \nonumber \\
&& \times\, \mbox{Tr} \left[ G(k,i\omega) {\bf v}_M
G(k,i\omega+i\Omega) {\bf v}_M \right]
\label{eq:Pidef}
\end{eqnarray}
where ${\bf v}_M \equiv {\bf v}_{MF}+{\bf v}_{M\Delta}$ is a
vector in coordinate space and a matrix in extended-Nambu space,
the Green's functions are dressed with disorder, the $\omega$-sum
is over fermionic Matsubara frequencies, the $k$-sum is restricted
to the first reduced Brillouin zone, the trace is over
extended-Nambu space, and $\beta=1/k_B T$.  We expand the $k$-sum
from the reduced Brillouin zone to the full (original) Brillouin
zone, which double-counts and therefore requires division by 2.
Since the summand is sharply peaked in the vicinity of the four
nodal collision points, we then replace the $k$-sum by four
integrals over local scaled coordinates, $p_1$ and $p_2$, defined
(in Sec.~\ref{ssec:spectrum}) about each of these points.
\begin{equation}
\sum_k^\prime \rightarrow \frac{1}{2}\sum_k \rightarrow
\frac{1}{2}\sum_{j=1}^4 \int \frac{d^2p}{(2\pi)^2 v_F v_\Delta}
\label{eq:sum2integrals}
\end{equation}
Making use of a spectral representation of the matrix Green's
function
\begin{equation}
G({\bf p},i\omega) = \int d\omega_1\,
\frac{-\frac{1}{\pi}\mbox{Im}\, G^R({\bf p},\omega_1)}{i\omega -
\omega_1}
\label{eq:specrep}
\end{equation}
Eq.~(\ref{eq:Pidef}) becomes
\begin{eqnarray}
\tensor{\Pi}_\kappa(i\Omega) &=& \frac{1}{2\pi^2 v_F v_\Delta}
\int \frac{d^2p}{(2\pi)^2} \int d\omega_1 d\omega_2 S(i\Omega)
\nonumber \\
&\times& \mbox{Tr} \left[ \sum_{j=1}^4 G_R^{''}({\bf p},\omega_1)
{\bf v}_M^{(j)} G_R^{''}({\bf p},\omega_2) {\bf v}_M^{(j)} \right]
\label{eq:PiofS}
\end{eqnarray}
where
\begin{equation}
S(i\Omega) = \frac{1}{\beta} \sum_{i\omega}
(i\omega+\frac{i\Omega}{2})^2 \frac{1}{i\omega - \omega_1}
\frac{1}{i\omega + i\Omega - \omega_2}
\label{eq:MatsuSum}
\end{equation}
and ${\bf v}_M^{(j)}$ is the value of ${\bf v}_M$ in the vicinity
of collision point $j$.  (Note that while the spectral
representation defined in Eq.~(\ref{eq:specrep}) is valid for the
case of real $\psi$ that we are considering, it would not be valid
if $\psi$, and therefore $H_k$, was complex.  The subtleties of
this are discussed in detail in the Appendix.)

Computing the Matsubara sum in Eq.~(\ref{eq:MatsuSum}) via contour
integration (see Refs.~\onlinecite{amb65} and \onlinecite{dur00}
for discussion of technical points), continuing $i\Omega
\rightarrow \Omega+i\delta$ to obtain the retarded function, and
taking the imaginary part, we find that
\begin{equation}
S^{''}_R(\Omega) = \pi (\omega_1 + \frac{\Omega}{2})^2
(n_F(\omega_1+\Omega) - n_F(\omega_1)) \delta(\omega_1 + \Omega -
\omega_2)
\label{eq:ImSret}
\end{equation}
where $n_F(x) = 1/(e^{\beta x} + 1)$ is the Fermi function the
double-prime indicates the imaginary part. Then taking the $\Omega
\rightarrow 0$ limit in Eq.~(\ref{eq:kubo}) yields an expression
for the thermal conductivity tensor
\begin{equation}
\frac{\tensor{\kappa}(T)}{T} = \frac{-1}{2\pi^2 v_F v_\Delta} \int
\! d\omega \left( \frac{\omega}{T} \right)^2 \frac{\partial
n_F}{\partial \omega} \int \frac{d^2p}{4\pi} \,\mbox{Tr}\,
\tensor{R}({\bf p},\omega)
\label{eq:kappaT}
\end{equation}
where
\begin{equation}
\tensor{R}({\bf p},\omega) = \sum_{j=1}^4 G_R^{''}({\bf
p},\omega_1) {\bf v}_M^{(j)} G_R^{''}({\bf p},\omega_2) {\bf
v}_M^{(j)}
\label{eq:Rdef}
\end{equation}
and taking the $T \rightarrow 0$ limit yields
\begin{equation}
\frac{\tensor{\kappa}_0}{T} = \frac{k_B^2}{6 v_F v_\Delta} \int
\frac{d^2p}{4\pi} \,\mbox{Tr}\, \tensor{R}({\bf p},0)
\label{eq:kappa0}
\end{equation}
Here we have used the fact that, for low $T$, $(\omega/T)^2
(-\partial n_F / \partial \omega)$ is sharply peaked at $\omega=0$
and
\begin{equation}
\int_{-\infty}^{\infty} \! d\omega \left( \frac{\omega}{T}
\right)^2 \left(-\frac{\partial n_F}{\partial \omega}\right) =
\frac{\pi^2 k_B^2}{3}
\label{eq:integ1}
\end{equation}
Noting that at each collision point, ${\bf v}_F$ and ${\bf
v}_\Delta$ point along the local $k_1$ and $k_2$ directions
respectively (as defined in Fig.~\ref{fig:BZcoordinates}) and
performing the sum over collision points in Eq.~(\ref{eq:Rdef}),
we find that
\begin{eqnarray}
\lefteqn{R_{ii}({\bf p},0) = 2v_F^2 G_R^{''} M_3^i G_R^{''} M_3^i
+ 2v_\Delta^2 G_R^{''} M_1^i G_R^{''} M_1^i} \nonumber \\
&& + 2\eta_i v_F v_\Delta (G_R^{''} M_3^i G_R^{''} M_1^i +
G_R^{''} M_1^i G_R^{''} M_3^i)
\end{eqnarray}
for $i=\{x,y\}$ while
\begin{equation}
R_{xy}({\bf p},0) = R_{yx}({\bf p},0) = 0
\label{eq:Rxy}
\end{equation}
where $\eta_i$, $M_3^i$, and $M_1^i$ are defined in
Eqs.~(\ref{eq:eta}) and (\ref{eq:M3M1}).  Plugging in the Green's
function from Eqs.~(\ref{eq:ImG} - \ref{eq:ImGden}) and taking the
trace over the $4 \times 4$ extended-Nambu space yields
\begin{equation}
\frac{\kappa_0^{ii}}{T} = \frac{\kappa_{00}}{T} \int
\frac{d^2p}{4\pi} \frac{N_1 + \eta_i N_2}{D}
\label{eq:kappa01}
\end{equation}
\begin{equation}
N_1 = 2A \left[ (A+B+\epsilon_1^2+\Delta_1^2)^2 +
(A+B+\epsilon_2^2+\Delta_2^2)^2 \right]
\label{eq:N1}
\end{equation}
\begin{equation}
N_2 = 4AB \left[ (\epsilon_1+\epsilon_2)^2 - (\Delta_1-\Delta_2)^2
\right]
\label{eq:N2}
\end{equation}
\begin{eqnarray}
D &=& \big[ (A+B+\epsilon_1^2 + \Delta_1^2)(A+B+\epsilon_2^2 +
\Delta_2^2) \nonumber \\
&-& B((\epsilon_1+\epsilon_2)^2 + (\Delta_1-\Delta_2)^2) \big]^2
\label{eq:D}
\end{eqnarray}
where
\begin{equation}
\frac{\kappa_{00}}{T} \equiv \frac{k_B^2}{3\hbar} \left(
\frac{v_F}{v_\Delta} + \frac{v_\Delta}{v_F} \right)
\label{eq:kappa0dSC}
\end{equation}
is the universal-limit thermal conductivity for a $d$-wave
superconductor (without charge order) and we have defined $A
\equiv \Gamma_0^2$ (our parameter of disorder) and $B \equiv
\psi^2$ (our parameter of charge order).  Inserting our
expressions for the $\epsilon$'s and $\Delta$'s from
Eq.~(\ref{eq:epsdelta}) and integrating over ${\bf p}$, we can
obtain the zero-temperature thermal conductivity as a function of
$\psi$, $\Gamma_0$, and $\alpha=v_F/v_\Delta$.

\section{Analytical Results: Clean, Isotropic Limit}
\label{sec:analytical}
In the clean ($A=\Gamma_0^2 \rightarrow 0$),
isotropic ($\alpha = v_F/v_\Delta=1$) limit, the integrals in
Eq.~(\ref{eq:kappa01}) can be performed analytically, providing us
with a closed-form expression for the thermal conductivity tensor
as a function of the charge density wave order parameter, $\psi$.
Selecting $\psi_c$ (the value of $\psi$ at which the nodes vanish)
as our energy unit, and for $\alpha=1$, Eq.~(\ref{eq:epsdelta})
becomes
\begin{eqnarray}
\epsilon_1 = p_1 + 1 && \Delta_1 = p_2 \nonumber \\
\epsilon_2 = p_2 + 1 && \Delta_2 = p_1
\label{eq:epsdeltaiso}
\end{eqnarray}
It is then useful to make a change of variables to
\begin{eqnarray}
q_1 &\equiv& p_1 - p_2 \nonumber \\
q_2 &\equiv& p_1 + p_2 + 1
\label{eq:qfromp}
\end{eqnarray}
such that
\begin{eqnarray}
\epsilon_1 = (q_1+q_2+1)/2 && \Delta_1 = (q_2-q_1-1)/2 \nonumber \\
\epsilon_2 = (q_2-q_1+1)/2 && \Delta_2 = (q_1+q_2-1)/2
\label{eq:epsdeltaq}
\end{eqnarray}
Note that this change of variables has a Jacobian of $1/2$, such
that $\int d^2p \rightarrow \frac{1}{2} \int d^2q$.  Therefore,
\begin{equation}
\frac{\kappa_0^{ii}}{\kappa_{00}} = \int \frac{d^2q}{8\pi}
\frac{N_1 + \eta_i N_2}{D}
\label{eq:kappa0iso}
\end{equation}
\begin{equation}
N_1 = 4A \left[ \left( A+B+\frac{q^2+1}{2} \right)^2 + q_1^2
\right] \label{eq:N1iso}
\end{equation}
\begin{equation}
N_2 = 4AB \left[ (q_2+1)^2 - q_1^2 \right]
\label{eq:N2iso}
\end{equation}
\begin{equation}
D = \left[ f + A(q^2+1+2B) + A^2 \right]^2
\label{eq:Diso}
\end{equation}
where
\begin{equation}
f = \frac{(q^2-1)^2}{4} + (q_2-B)^2
\label{eq:fiso}
\end{equation}
In the $A \rightarrow 0$ limit, the numerator vanishes, so
contributions to the integral come only from the vicinity of
points in $q$-space where the denominator vanishes as well, which
requires $f=0$.  It is clear from Eq.~(\ref{eq:fiso}) that $f$ is
only equal to zero when $q=1$ and $q_2=B$, the intersection of a
unit circle about the origin and a horizonal line at $q_2=B$.

For $B>1$, there is no intersection, so the integral is zero.
This is quite physical, since for $B>1$, $\psi > \psi_c$ and the
energy spectrum is gapped.  Thus, in the clean, zero-temperature
limit, there are no quasiparticles to transport heat and the
thermal conductivity is zero.

\begin{figure}
\centerline{\resizebox{3.25in}{!}{\includegraphics{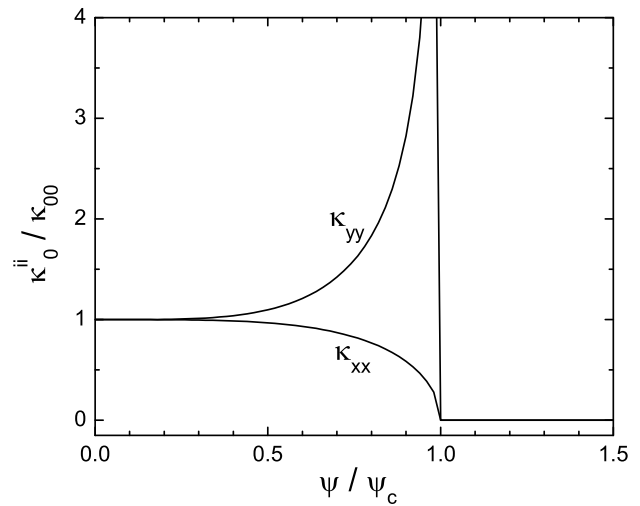}}}
\caption{Calculated zero-temperature thermal conductivity tensor
in the clean ($\Gamma_0 \rightarrow 0$), isotropic ($v_F =
v_\Delta$) limit.  We plot $\kappa_0^{xx}$ and $\kappa_0^{yy}$ as
functions of the charge density wave order parameter, $\psi$, from
the closed-form expressions in Eqs.~(\ref{eq:kappa0xx}) and
(\ref{eq:kappa0yy}).  As $\psi$ approaches $\psi_c$, the value
beyond which the quasiparticle spectrum becomes gapped,
$\kappa_0^{xx}$ vanishes continuously while $\kappa_0^{yy}$
diverges before dropping to zero.}
\label{fig:kappa0clean}
\end{figure}

For $B<1$, the circle and line intersect at two points, ${\bf
q}_n=(\pm\sqrt{1-B^2},B)$.  These points are precisely the node
and ghost-node of the energy spectrum, which will collide when
$\psi$ reaches $\psi_c$.  For vanishing $A$, terms in $N_1$,
$N_2$, and $D$ that are higher than first order in $A$ can be
safely neglected and terms first order in $A$ can be replaced by
their values at ${\bf q}={\bf q}_n$.  Doing so, we find that
\begin{equation}
\frac{\kappa_0^{ii}}{\kappa_{00}} = (1+\eta_i B^2)8(1+B)I_1
\label{eq:kappa0clean}
\end{equation}
where
\begin{equation}
I_1 \equiv \int \frac{d^2q}{8\pi} \frac{A}{\left[ f + 2A(1+B)
\right]^2}
\label{eq:I1def}
\end{equation}
and $f$ is the function of ${\bf q}$ given in Eq.~(\ref{eq:fiso}).
Changing variables to
\begin{eqnarray}
x_1 &\equiv& q_1 - 1 = x\cos\theta \nonumber \\
x_2 &\equiv& q_2 = x\sin\theta
\label{eq:xfromq}
\end{eqnarray}
we see that
\begin{eqnarray}
f &=& x^4/4 + B^2 + x^2 + x^3\cos\theta - 2Bx\sin\theta \nonumber \\
&=& \frac{x^2}{h^2} \left[ 1 + h^2 + 2h(\cos\theta \cos\theta_0 -
\sin\theta \sin\theta_0) \right] \nonumber \\
&=& \frac{x^2}{h^2} \left[ 1 + h^2 + 2h\cos(\theta+\theta_0)
\right]
\label{eq:fx}
\end{eqnarray}
where
\begin{equation}
h \equiv \frac{x}{\sqrt{x^4/4 + B^2}} \;\;\; \mbox{and} \;\;\;
\tan\theta_0 \equiv \frac{2B}{x^2} \label{eq:hdef}
\end{equation}
Then plugging $f$ into Eq.~(\ref{eq:I1def}), shifting $\theta
\rightarrow \theta - \theta_0 + \pi$, and defining $\gamma \equiv
2(1+B)h^2/x^2$, we find that
\begin{eqnarray}
I_1 &=& \frac{1}{8\pi} \int_0^\infty \!\! dx\, x \int_{-\pi}^\pi
\!\! d\theta \frac{h^4}{x^4} \frac{A}{\left[ 1 + h^2 -
2h\cos\theta + A\gamma \right]^2} \nonumber \\
&=& \frac{1}{8\pi(1+B)} \int_0^\infty \frac{dx}{x} h^2 I_2
\label{eq:I1ofI2}
\end{eqnarray}
where
\begin{equation}
I_2 \equiv \int_0^\pi \!\! d\theta \frac{A\gamma}{\left[ 1 + h^2 +
A\gamma - 2h\cos\theta \right]^2}
\label{eq:I2def}
\end{equation}
This integral over $\theta$ is standard and easily evaluated via
integration table \cite{gra94}.  Doing so yields
\begin{equation}
I_2 = 2\pi \frac{1+h^2}{(1+h)^3} \, D(h-1,A\gamma)
\label{eq:I2ofD}
\end{equation}
where
\begin{equation}
D(u,\Gamma) \equiv \frac{\Gamma^2/2}{(u^2+\Gamma^2)^{3/2}}
\label{eq:Ddef}
\end{equation}
Since $\gamma$ is finite for all $x$, $A\gamma$ vanishes as $A
\rightarrow 0$.  Therefore, noting that
\begin{equation}
\lim_{\Gamma \rightarrow 0} D(u,\Gamma) = \Big\{
\begin{array}{c}
0 \;\;\mbox{for}\;\; u \neq 0 \\
\infty \;\;\mbox{for}\;\; u=0
\end{array}
\label{eq:Dlimits}
\end{equation}
and
\begin{equation}
\int_{-\infty}^\infty \!\! du\, D(u,\Gamma) = 1
\label{eq:Dint}
\end{equation}
we see that $D(u,\Gamma \rightarrow 0)$ is a representation of the
Dirac delta function.  Hence,
\begin{equation}
I_2 = \frac{\pi}{2} \delta(h-1)
\label{eq:I2final}
\end{equation}
and
\begin{equation}
I_1 = \frac{1}{16(1+B)} \int_0^\infty \!\! dx
\,\frac{x}{x^4/4+B^2}\, \delta(h-1)
\label{eq:I1h}
\end{equation}
Noting that $\delta(h-1)=2\delta(h^2-1)$ (since $h>1$) and letting
$u \equiv x^2/2$, this becomes
\begin{equation}
I_1 = \frac{1}{8(1+B)} \int_0^\infty \frac{du}{u^2+B^2} \,\delta\!
\left( \frac{2u}{u^2+B^2}-1 \right)
\label{eq:I1u}
\end{equation}
For $B>1$, the argument of the delta function is never zero, so
$I_1=0$, as expected. For $B<1$, the argument is zero at two
points, $u=u_\pm=1 \pm \sqrt{1-B^2}$, so after a bit of
delta-function gymnastics, we find that
\begin{eqnarray}
I_1 &=& \frac{1}{16(1+B)} \frac{1}{\sqrt{1-B^2}} \int_0^\infty
\!\! du \left[ \delta(u-u_-) + \delta(u-u_+) \right] \nonumber \\
&=& \frac{1}{8(1+B)} \frac{\Theta(1-B)}{\sqrt{1-B^2}}
\label{eq:I1final}
\end{eqnarray}
where $\Theta(x)$ is the Heaviside step function.  Finally,
plugging back into Eq.~(\ref{eq:kappa0clean}), we obtain a very
simple, closed-form expression for the zero-temperature thermal
conductivity tensor in the clean, isotropic limit.
\begin{equation}
\frac{\kappa_0^{xx}}{\kappa_{00}} = \sqrt{1-(\psi/\psi_c)^4}
\,\,\Theta(\psi_c-\psi)
\label{eq:kappa0xx}
\end{equation}

\begin{equation}
\frac{\kappa_0^{yy}}{\kappa_{00}} =
\frac{1+(\psi/\psi_c)^4}{\sqrt{1-(\psi/\psi_c)^4}}
\,\,\Theta(\psi_c-\psi)
\label{eq:kappa0yy}
\end{equation}
\begin{equation}
\kappa_0^{xy} = \kappa_0^{yx} = 0
\label{eq:kappa0xy}
\end{equation}
These results are plotted in Fig.~\ref{fig:kappa0clean}.  For
$\psi=0$, we recover the universal-limit thermal conductivity of a
$d$-wave superconductor \cite{dur00} (see
Eq.~(\ref{eq:kappa0dSC})). And for $\psi > \psi_c$, as expected,
the thermal conductivity vanishes since the system has become
gapped and there are no quasiparticles to transport the heat.  For
$\psi$ between zero and $\psi_c$, thermal transport in the $x$ and
$y$ directions differ, which makes sense as square symmetry has
been explicitly broken by the charge density wave oriented in the
$x$-direction.  Parallel to the CDW wave vector, thermal
conductivity in the $x$-direction decreases monotonically with
$\psi$, vanishing continuously at $\psi_c$.  Perpendicular to the
CDW wave vector, thermal conductivity in the $y$-direction
increases with $\psi$, exhibiting a square-root divergence before
vanishing abruptly at $\psi_c$.  This divergence, a consequence of
the clean limit, is replaced by a peak in $\kappa_0^{yy}$ when
nonzero disorder is considered, as will be shown in the next
section.

\section{Numerical Results}
\label{sec:numerical}
\begin{figure}
\centerline{\resizebox{4in}{!}{\includegraphics{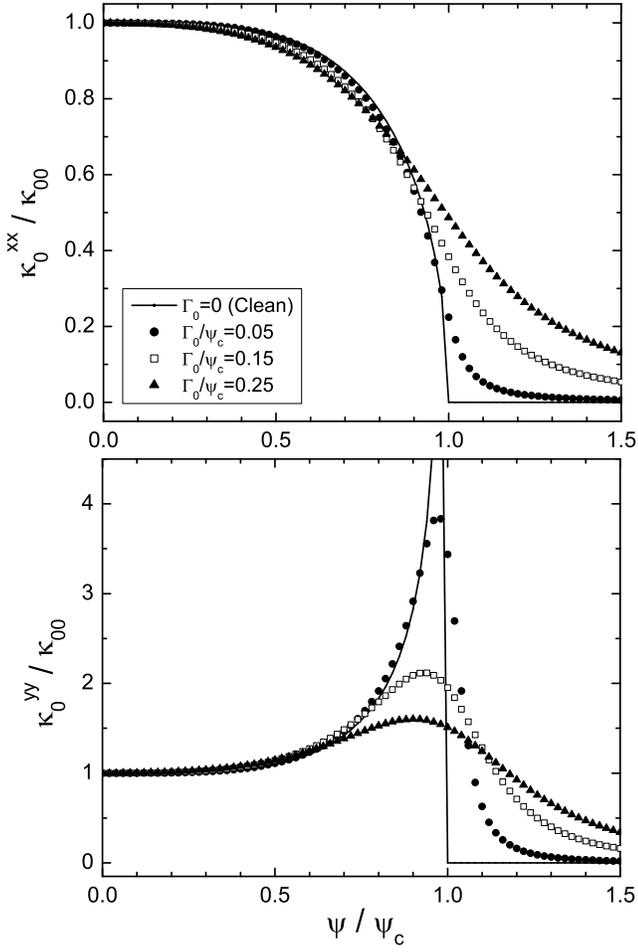}}}
\caption{Disorder dependence of calculated zero-temperature
thermal conductivity tensor.  We plot $\kappa_0^{xx}$ (upper
panel) and $\kappa_0^{yy}$ (lower panel) as functions of charge
density wave order parameter, $\psi$, for several values of the
disorder parameter, $\Gamma_0$.  In all cases, $\alpha = v_F /
v_\Delta = 1$. Included for comparison is the clean limit result
(solid lines). Note that disorder smoothes the transition to zero
thermal conductivity that results from the gapping of the energy
spectrum at $\psi = \psi_c$.  The divergence of $\kappa_0^{yy}$
seen in the clean case is replaced by a peak that is diminished
and broadened with increasing disorder.}
\label{fig:kappa0Gamma0}
\end{figure}

\begin{figure}
\centerline{\resizebox{4in}{!}{\includegraphics{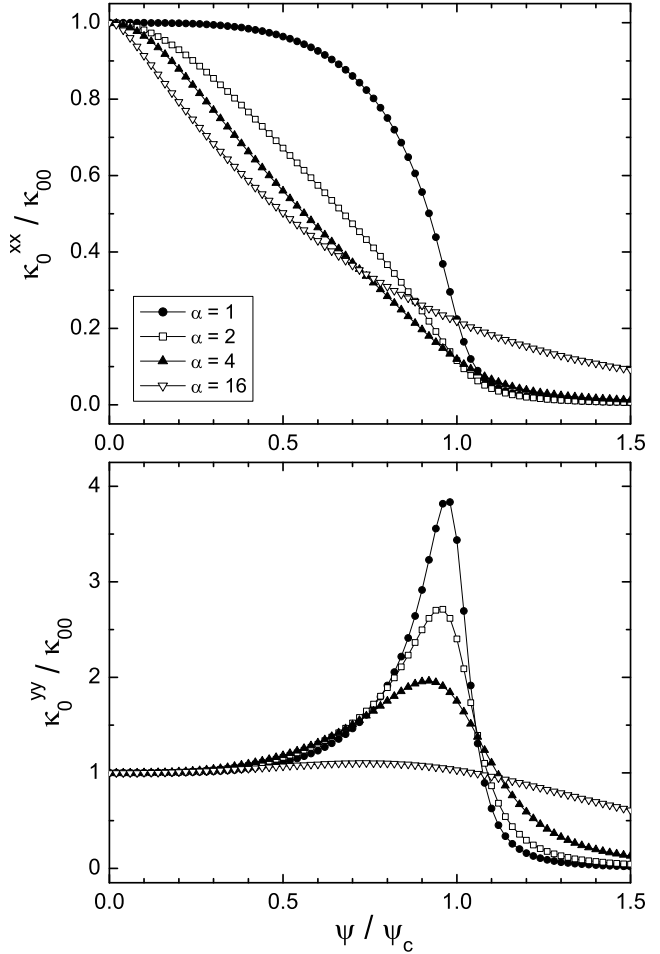}}}
\caption{Nodal anisotropy dependence of calculated
zero-temperature thermal conductivity tensor.  For fixed disorder
($\Gamma_0 = 0.05 \psi_c$), we plot $\kappa_0^{xx}$ (upper panel)
and $\kappa_0^{yy}$ (lower panel) as functions of charge density
wave order parameter, $\psi$, for several values of $\alpha =
v_F/v_\Delta$.  Lines connecting the data points are guides to the
eye.  Note that the nodal transition at $\psi = \psi_c$ is
smoothed out by increasing velocity anisotropy in a manner similar
to, but distinct from, the effect of disorder.}
\label{fig:kappa0alpha}
\end{figure}

For the general case of nonzero disorder
($\Gamma_0 \neq 0$) and/or anisotropic Dirac nodes ($\alpha = v_F
/ v_\Delta \neq 1$), the $p$-space integration in
Eq.~(\ref{eq:kappa01}) is more complicated, but can be computed
numerically.  Doing so, we calculated the zero-temperature thermal
conductivity tensor as a function of charge density wave order
parameter, $\psi$, our parameter of disorder, $\Gamma_0$, and the
anisotropy of the Dirac nodes, $\alpha=v_F/v_\Delta$. Results are
plotted in Figs.~\ref{fig:kappa0Gamma0} and \ref{fig:kappa0alpha}.

Fig.~\ref{fig:kappa0Gamma0} shows $\kappa_0^{xx}$ and
$\kappa_0^{yy}$ as functions of $\psi$ for several values of
$\Gamma_0$ and $\alpha=1$.  The clean-limit results calculated in
Sec.~\ref{sec:analytical} are included (solid lines) for
comparison. Note that as disorder increases, the transition to
zero thermal conductivity at the nodal collision point
($\psi=\psi_c$) gets rounded out, and the peak in $\kappa_0^{yy}$
just prior to the collision point is diminished and broadened.
Essentially, and not unexpectedly, disorder blurs the nodal
collision, smoothing out the sharp transition seen in the clean
case.

The $\Gamma_0=0.05 \psi_c$ results are reproduced in
Fig.~\ref{fig:kappa0alpha}, along with plots of $\kappa_0^{xx}$
and $\kappa_0^{yy}$ versus $\psi$ for larger values of $\alpha$.
For constant disorder, increasing $\alpha$ changes the shape of
the $\kappa_0^{xx}$ curve and diminishes and broadens the peak in
$\kappa_0^{yy}$.  The effect of increased nodal anisotropy is
similar to, but distinct from, that of disorder, further smoothing
the transition to zero thermal conductivity that occurs abruptly
at $\psi=\psi_c$ in the clean, isotropic case.

\section{Conclusions}
\label{sec:conclusions}
The coexistence of $d$-wave
superconductivity with charge order of sufficient magnitude can
have a significant effect on the energy spectrum of the Bogoliubov
quasiparticles and the transport of heat by those quasiparticles
at low temperatures. In this paper, we have considered a
particularly simple form of charge order, a conventional $s$-wave
charge density wave of wave vector ${\bf Q}=(\pi/a,0)$, the
magnitude of which is characterized by a real, $k$-independent
order parameter, $\psi$. The charge order halves the Brillouin
zone, and as a function of $\psi$, the four nodes of the
quasiparticle energy spectrum move in $k$-space, approaching the
reduced Brillouin zone edge.  When $\psi$ reaches $\psi_c$, equal
to the Fermi velocity times the $k$-space distance from the
original node location to the $(\pi/2,\pi/2)$ point, the nodes
reach the reduced Brillouin zone edge and collide with their
counterparts in the second reduced Brillouin zone.  Beyond this
point, the nodes vanish and the quasiparticle energy spectrum is
fully gapped.

We have used a linear response Kubo formula approach to calculate
the zero temperature limit of the thermal conductivity tensor for
this system.  Working within an extended-Nambu basis (particle,
hole, particle shifted by ${\bf Q}$, hole shifted by ${\bf Q}$),
we constructed a $4 \times 4$ matrix Hamiltonian, Green's
function, and thermal current operator.  We then used the
Matsubara technique to evaluate the bare-bubble thermal
current-current correlator, neglecting vertex corrections and
including disorder in the self-energy via a single broadening
parameter, $\Gamma_0$.  From this we calculated $\kappa^{xx}/T$
and $\kappa^{yy}/T$, in the limit of zero temperature, as a
function of $\psi$, $\Gamma_0$, and the nodal anisotropy $\alpha =
v_F / v_\Delta$.

In the clean ($\Gamma_0 \rightarrow 0$), isotropic ($v_F =
v_\Delta$) limit, our calculations yield a closed-form solution
for the thermal conductivity tensor (plotted in
Fig.~\ref{fig:kappa0clean})
\begin{equation}
\frac{\kappa_0^{xx}}{\kappa_{00}} = \sqrt{1-(\psi/\psi_c)^4}
\,\,\Theta(\psi_c-\psi)
\label{eq:kappa0xxagain}
\end{equation}
\begin{equation}
\frac{\kappa_0^{yy}}{\kappa_{00}} =
\frac{1+(\psi/\psi_c)^4}{\sqrt{1-(\psi/\psi_c)^4}}
\,\,\Theta(\psi_c-\psi)
\label{eq:kappa0yyagain}
\end{equation}
\begin{equation}
\kappa_0^{xy} = \kappa_0^{yx} = 0
\label{eq:kappa0xyagain}
\end{equation}
where
\begin{equation}
\frac{\kappa_{00}}{T} \equiv \frac{k_B^2}{3\hbar} \left(
\frac{v_F}{v_\Delta} + \frac{v_\Delta}{v_F} \right)
\label{eq:kappa0dSCagain}
\end{equation}
is the zero-temperature thermal conductivity for a $d$-wave
superconductor with no charge order.  As expected, the thermal
conductivity takes the pure $d$-wave superconductor value for
$\psi=0$ and drops to zero for $\psi > \psi_c$, where the
quasiparticle energy spectrum has become fully gapped.  For
intermediate values of $\psi$, $\kappa_0^{xx}$ and $\kappa_0^{yy}$
differ, as square symmetry has been broken by the charge density
wave.  For transport in the direction of the charge density wave
vector, $\kappa_0^{xx}$ vanishes continuously as $\psi$ approaches
$\psi_c$.  By contrast, for transport perpendicular to the charge
density wave vector, $\kappa_0^{yy}$ diverges before dropping
abruptly to zero at $\psi_c$.  This divergence is a consequence of
the clean limit and is replaced by a finite peak in the presence
of disorder.

For the more complicated case of nonzero disorder ($\Gamma_0 \neq
0$) and/or anisotropic nodes ($v_F \neq v_\Delta$), we have
obtained results via a numerical calculation.  We find that
disorder smoothes out the transition to zero thermal conductivity
across the nodal collision (see Fig.~\ref{fig:kappa0Gamma0}).  The
clean-limit divergence in $\kappa_0^{yy}$ just before the
transition is replaced by a peak which broadens and decreases in
amplitude with increasing disorder.  The abrupt drop in the
clean-limit $\kappa_0^{xx}$ is similarly broadened.  Essentially,
the disorder-broadening of the quasiparticle spectral function
averages over what was, in the clean limit, a sharp transition
from gapless to gapped quasiparticles.  We find that increased
nodal anisotropy has a similar effect, amplifying the disorder
effect and thereby further broadening the features in the
$\psi$-dependence of the thermal conductivity (see
Fig.~\ref{fig:kappa0alpha}).  And the fact that disorder has an
effect indicates that the low-temperature thermal conductivity is
no longer universal (disorder-independent) in the presence of
charge order, which is in line with the results of recent
measurements \cite{and04,sun05,sun06,haw03} of low-temperature
thermal transport in the underdoped cuprates.

In these calculations, we have enjoyed the theorist's luxury of
being able to turn on, by hand, a charge density wave to coexist
with the $d$-wave superconductivity.  The experimenter does not
have direct access to such a knob.  However, in the $d$-wave
superconducting state of the cuprates, charge order does appear to
be enhanced with underdoping.  Hence the features of the
$\psi$-dependent thermal conductivity curves calculated herein
should serve as signatures for the underdoping-dependence of
thermal conductivity measured in the underdoped cuprates.  Of
course, most dramatic would be the observation of the nodal
collision beyond which the low-temperature thermal conductivity
drops to zero.  However, even if the amplitude of charge order is
insufficient to reach the nodal collision, these results should
provide insight to the approach to the transition.

A sequel to this work, exploring the effects of a more elaborate
model of disorder, as well as the contribution of vertex
corrections, is in preparation \cite{sch08}.  Future work will
also examine the effect of different types of charge order (beyond
the conventional $s$-wave case considered here) of different wave
vector (beyond the unit-cell-doubling ${\bf Q} = (\pi/a,0)$ case
considered here) and of multiple wave vectors (like the
checkerboard charge order observed in some cuprates
\cite{han04,wis08}).

\begin{acknowledgments}
We are grateful to S. M. Girvin, A. Abanov, and P. Schiff for very
helpful discussions.  This work was supported by NSF Grants No.\
DMR-0605919 (A.C.D.) and No.\ DMR-0757145 (S.S.).
\end{acknowledgments}

\appendix*

\section{Subtleties of the Spectral Representation}
\label{app:specrep}
In the calculations described in this paper, we
have made use of the $4 \times 4$ extended-Nambu basis of
Eq.~(\ref{eq:basis}) for the Hamiltonian and Green's functions.
This choice of basis provides a compact realization of the
Hamiltonian and is quite convenient in many respects.  However,
use of a matrix Green's function does introduce some subtleties
regarding the spectral representation, and we would like to
address those here.

All of our results could have been obtained by diagonalizing the
Hamiltonian from the outset and working with the diagonalized
Green's function
\begin{equation}
G_D(i\omega) = U^\dagger G(i\omega) U
\label{eq:Gdiag}
\end{equation}
with diagonal matrix elements
\begin{equation}
\left[ G_D(i\omega) \right]_{nn} = \frac{1}{i\omega - E_k^{(n)} -
\Sigma^{(n)}(i\omega)}
\label{eq:Gdiagelements}
\end{equation}
where the $E_k^{(n)}$ are the eigenvalues of $H_k$, the
$\Sigma^{(n)}$ are the corresponding self-energies, and the
eigenvectors define the columns of unitary transformation matrix
$U$. In the diagonal basis, it is quite valid to define a spectral
representation for the Green's function
\begin{equation}
G_D(i\omega) = \int d\omega_1
\frac{A_D(\omega_1)}{i\omega-\omega_1}
\label{eq:Gdiagspecrep}
\end{equation}
where
\begin{equation}
A_D(\omega) \equiv \frac{i}{2\pi} \left[ G_D^R(\omega) -
G_D^A(\omega) \right] = -\frac{1}{\pi} \mbox{Im} G_D^R(\omega)
\label{eq:Adiag}
\end{equation}
Note that the second equality follows from the fact that the
retarded diagonal Green's function, $G_D^R(\omega) \equiv
G_D(i\omega \rightarrow \omega+i\delta)$, is the complex conjugate
of the advanced diagonal Green's function, $G_D^A(\omega) \equiv
G_D(i\omega \rightarrow \omega-i\delta)$, which is clear from
Eq.~(\ref{eq:Gdiagelements}).

The non-diagonal matrix Green's function can therefore be
expressed as
\begin{equation}
G(i\omega) = U G_D(i\omega) U^\dagger = \int d\omega_1
\frac{-\frac{1}{\pi} U \mbox{Im} G_D^R(\omega_1)
U^\dagger}{i\omega - \omega_1}
\label{eq:GfromGDspecrep}
\end{equation}
which is not equivalent to the right-hand side of
Eq.~(\ref{eq:specrep})
\begin{equation}
\int d\omega_1 \frac{-\frac{1}{\pi} \mbox{Im}
G^R(\omega_1)}{i\omega-\omega_1} = \int d\omega_1
\frac{-\frac{1}{\pi} \mbox{Im} \left[ U G_D^R(\omega_1) U^\dagger
\right]}{i\omega-\omega_1}
\label{eq:Gspecrep}
\end{equation}
unless the diagonalization transformation commutes with taking the
imaginary part.  Equivalently, note that it is valid to define a
non-diagonal matrix spectral function, $A(\omega) \equiv U
A_D(\omega) U^\dagger$, such that
\begin{equation}
G(i\omega) = \int d\omega_1 \frac{A(\omega_1)}{i\omega-\omega_1}
\label{eq:Gnondiagspecrep}
\end{equation}
but
\begin{equation}
A(\omega) \equiv \frac{i}{2\pi} \left[ G^R(\omega) - G^A(\omega)
\right]
\label{eq:Anondiag}
\end{equation}
will not be equal to $-\mbox{Im}G^R/\pi$ (or even have to be real)
unless the non-diagonal retarded Green's function, $G^R(\omega)
\equiv G(i\omega \rightarrow \omega+i\delta)$, is the complex
conjugate of the non-diagonal advanced Green's function,
$G^A(\omega) \equiv G(i\omega \rightarrow \omega-i\delta)$.

For the case of real $\psi$ that we consider in this paper, $H_k$
and therefore $U$ are real, so diagonalization does commute with
taking the imaginary part, $G^R$ is the complex conjugate of
$G^A$, and Eq.~(\ref{eq:specrep}) is valid. But this is not
generically the case for complex $\psi$.

\end{document}